\documentclass[12pt,preprint]{aastex}

\newcommand{\hone}{H~{\footnotesize{I}}}  	
\newcommand{\halpha}{H{${\alpha}$}}	
\newcommand{\oone}{[O{\footnotesize{I]}}}	
\newcommand{\nenine}{Ne~{\footnotesize{IX}}}    
\newcommand{\neten}{Ne~{\footnotesize{X}}}      
\newcommand{\mageleven}{Mg~{\footnotesize{XI}}} 
\newcommand{\magtwelve}{Mg~{\footnotesize{XII}}} 
\newcommand{\sithirteen}{Si~{\footnotesize{XIII}}} 
\newcommand{\sifourteen}{Si~{\footnotesize{XIV}}} 
\newcommand{\stwo}{[S{\footnotesize{II]}}}	
\newcommand{\sfifteen}{S~{\footnotesize{XV}}}   
\newcommand{\ironseventeen}{Fe~{\footnotesize{XVII}}} 

\shorttitle{W44}
\shortauthors{Shelton, Kuntz, and Petre}
\received{2002 August 5}
\begin{document}


\title{{\it{Chandra}} Observations and Models of the Mixed Morphology 
Supernova Remnant W44: Global Trends}

\author{R. L. Shelton\altaffilmark{1}\altaffilmark{2}, 
K. D. Kuntz\altaffilmark{3},
R. Petre\altaffilmark{4}}
\affil{The Department of Physics and Astronomy and the 
Center for Simulational Physics at the University of Georgia,
Athens, GA 30602}
\affil{The Department of Physics at the Univeristy 
University of Maryland at Baltimore County}
\affil{NASA's Goddard Space Flight Center}
\email{rls@hal.physast.uga.edu}
\email{kuntz@milkyway.gsfc.nasa.gov}
\email{rob@milkyway.gsfc.nasa.gov}

\begin{abstract}

We report on the {\it{Chandra}} observations of the archetypical mixed
morphology (or thermal composite) supernova remnant, W44.
As with other mixed morphology remnants, W44's projected center is
bright in thermal X-rays.  It has an obvious radio shell, but
no discernable X-ray shell.
In addition, X-ray bright
'knots' dot W44's image.  The spectral analysis of the {\it{Chandra}}
data show that the remnant's hot, bright projected center is 
metal-rich and that the bright knots are regions of comparatively
elevated elemental
abundances.  Neon is among the affected elements, suggesting that
ejecta contributes to the abundance trends.  Furthermore, some of
the emitting iron atoms appear to be underionized with respect to
the other ions, providing the first potential X-ray evidence for 
dust destruction in a supernova remnant.  We use the {\it{Chandra}}
data to test the following explanations for W44's X-ray bright
center:  
1.) entropy mixing due to bulk mixing or thermal conduction,
2.) evaporation of swept up clouds, and
3.) a metallicity gradient, possibly due to dust destruction and ejecta 
enrichment.
In these tests, we assume that the remnant has evolved beyond the 
adiabatic evolutionary stage, which explains the X-ray dimness of the shell.
The entropy mixed model
spectrum was tested against the {\it{Chandra}} spectrum for the remnant's
projected center and found to be a good match.
The evaporating clouds model was constrained by the finding that 
the ionization parameters of the bright knots are similar to those
of the surrounding regions.
%
While both the
entropy mixed and the evaporating clouds models are known to predict centrally
bright X-ray morphologies, their predictions fall short of the 
observed brightness gradient.  The resulting brightness gap can be largely
filled in by emission from the extra metals in and near the remnant's
projected center.  The preponderance of evidence (including that
drawn from other studies) suggests
that W44's remarkable morphology can be attributed to dust destruction
and ejecta enrichment within an entropy mixed, adiabatic phase
supernova remnant.  The {\it{Chandra}} data prompts a new question
-- by what astrophysical mechanisms are the metals distributed so 
inhomogeneously in the supernova remnant.

\end{abstract}


\section{Introduction}

	Of the middled aged remnants, W44 is remarkably well studied.
X-ray observations of this supernova remnant (SNR)
reveal centrally concentrated, thermal X-ray emission
(Smith et al. (1985), Jones et al. (1993),
Rho et al. (1994), and Harrus et al. (1997)), while
radio observations reveal a filamentary,
edge brightened, elongated, synchrotron emission shell
(see the beautiful radio-frequency images in 
Jones, Smith, and Angellini's (1993)).
Because W44 has both a thermal X-ray bright center
and a radio synchrotron shell, it is the epitome
of 'thermal composite' or 'mixed morphology' supernova remnants.
Such remnants form a puzzling class.  No simple 
models of supernova remnants evolving in a homogeneous ambient medium
can explain the tremendous fluxes of 
thermal X-rays emitted by the centers of these remnants.
For this reason, remnants like W44 have become specimens with which to
examine plasma physics, remnant evolution, and the inhomogeneous
interstellar medium (ISM).
Here, we follow up on such studies by examining the
new {\it{Chandra}} ACIS data.

Section 2 presents
a more comprehensive description of W44, including its environment,
cool shell, pulsar, and X-ray 
characteristics.
During the W44's long history as a specimen for examination,
many interpretations have been suggested.  These are presented
in Section 3.
The {\it{Chandra}} images and spectra
are presented, examined, and compared with models in Section 4.
Section 5 summarizes the results. 

\section{Summary of W44 Explorations}

\subsection{Setting}

W44, also called 3C~392, lies in the Galactic plane,
at $l = 34.7^{\rm{o}}$, $b = -0.4^{\rm{o}}$
(R.A. $= 18^{\rm{h}} 56^{\rm{m}} 05^{\rm{s}}$, dec =
$01^{\rm{o}} 23^{\rm{'}} 28^{\rm{''}}$ in (J2000) coordinates).
According to Cox et al.'s (1999) calculations from the \citet{caswell_etal}
21 cm data and the
\citet{clemens}
Galactic rotation curves,
W44 has a distance of 2.5 to 2.6 kpc.
When viewed in projection, W44  appears to reside within
a forest of dense clouds
\citep{sato,dame_etal}.
Given W44's proximity to the molecular clouds,
it is not surprising that the environmental interstellar matter 
is thought to be relatively dense
($\sim 6$ atoms cm$^{-3}$, \citet{cox_etal}).

\subsection{Shell Observed in Radio-Continuum, \hone, \halpha,
and \stwo}

	The first reports of W44's
radio continuum emission came in 1958 \citep{westerhout}.
In the course of later analyses, W44 was identified as a
supernova remnant and labeled as
``shell type.''
Although the literal interpretation of this term brings to mind
a bright ring of emission,
W44's radio-continuum shell is well developed only in the northeast
quadrant (assuming equatorial coordinates,
see figures in \citet{jones_etal,giacani_etal} and this paper's
Figure~\ref{fig:xray_radio}).
In contrast,
W44's radio continuum flux gradient is more gradual and its
silhouette is more ragged in the other quadrants.
\citet{cox_etal} interpreted the radio continuum morphology as
that of a partially formed compressed shell.  The shell is mature
in the aft northeast section, but nowhere else (see Figure 1 of
\citet{cox_etal}).
The overall dimensions of this remnant are 25' $\times$ 35'
(from \citet{giacani_etal}'s 1442.5 MHz image), the shape
is elliptical or pear-like, and the semi-major
axis is oriented northwest-southeast.
The radio-continuum emission's spectral index is
$\sim\-0.3$ to $-0.4$ \citep{kovalenko_etal,kassim,giacani_etal},
indicative of radio synchrotron emission from electrons
in the shocked and
swept up gas at the remnant's periphery.

	Part of a shell has also been observed via tracers of neutral
material.
\citet{koo_heiles}'s velocity resolved \hone\ 21 cm emission
data for W44 exhibit the signature of the back (receding) side of a
shell expanding at $\sim150$ km sec$^{-1}$,
leading \citet{cox_etal} to think that the shell is incomplete
and more like a cap than a sphere.
(For an alternate explanation, see \citet{koo_heiles}.)
%
Furthermore, the published optical data suggest that W44 has
partial shells of \halpha\ and \stwo.
In \citet{giacani_etal}'s images, \stwo\ outlines W44's
north and west quadrants, while \halpha\ outlines shorter
segments in the north and west.  In some regions the
optical images correlate well with the radio emission
images, as expected in gas compressed by radiative shocks.
These indications of the presence of a cool, compressed
shell, work in concert with the current interpretations
for W44's X-ray morphology (see section 3.1).
Another intersection between the optical interpretations and
the X-ray interpretations (see Section 3.2) was noted by \citet{rho_etal}.
Noting that the remnant's face (when viewed in projection)
emits \stwo\ and \halpha,
\citet{rho_etal} suggested that if
the interior of the remnant contains evaporating clouds,
then the optical emission might arise from the evaporated material.
However, the evaporation zones would also be
expected to emit soft X-rays and there is little correlation
between the soft X-ray images and
the optical images.

The remnant has been successfully searched for
OH 1720.53 MHz maser emission \citep{claussen_etal}
and \oone\ emission \citep{reach_rho}.
These researchers suggest that these signals result from collisions between
the SNR and molecular clouds,
while \citet{cox_etal} suggest that they originate in the denser parts
of the remnant's shell.

\subsection{Pulsar Establishes W44's Age}

In 1991, Wolszczan, Cordes \& Dewey reported
a 20,000 year old
pulsar near W44's projected center and at about the same
distance. This pulsar, called PSR B1853+01, and attributed to W44,
appears to have a wind nebula that emits
radio frequency and X-ray synchrotron photons
\citep{harrus_etal_nebula,petre_etal,frail_giacani_etal}.
As seen in \citet{petre_etal}'s Figure 2 (made with this
{\it{Chandra}} data), the spatial extent of the nebula's
nonthermal X-ray emission region is only half an arcminute.
Neither the pulsar nor its wind nebula are thought
to affect the supernova remnant noticeably.

\subsection{X-ray Observations Reveal W44's Unusual Characteristics}

	Prior to the {\it{Chandra}} era, W44 had been observed by
the {\it{Astronomical Netherlands Satellite}} ({\it{ANS}}),
{\it{Einstein}},
the {\it{European X-ray Observatory Satellite}} ({\it{EXOSAT}}),
the {\it{Roentgen Satellite}} ({\it{ROSAT}}), and
the {\it{Advanced Satellite for Cosmology and Astrophysics}}
({\it{ASCA}}) \citep{gronenschild_etal,szymkowiak,
smith_etal,jones_etal,rho_etal,
harrus_etal_nebula,harrus_etal,shelton_etal}.
From the imaging
observations, we know that the
entire projected area of W44 (as defined by the radio continuum 
emission) is bright in X-ray emission.
The X-ray surface brightness peaks near the
projected center of the remnant and drops nonmonotonically
toward the edge of the remnant.
Also, W44 has several
bright diffuse regions, but no limb or edge brightening.

W44's spectra reveal
emission lines, the handiwork of thermal processes.
Least squares fitting between the previously observed spectra and 
collisional ionization equilibrium spectral
models favored temperatures from $\sim4$ to $\sim10 \times 10^6$~K
\citep{smith_etal,jones_etal,rho_etal,harrus_etal_nebula}.
Fits to the {\it{ROSAT}}
spectra for various regions showed that the
best-fit temperature varies little across the remnant
\citep{rho_etal}.
Furthermore, the plasma within the remnant's central region appears 
to be near collisional ionization equilibrium;
the ionization timescale 
of $2.0^{+4.3}_{-0.7} \times 10^{11}$ s cm$^{-3}$ and the temperature
of $\sim1.02 \pm 0.16 \times 10^{7}$~K, found from
the combined {\it{ROSAT}}, {\it{Einstein}}, and
{\it{Ginga}} data \citep{harrus_etal}
place the plasma near the collisional
ionization equilibrium state on \citet{vedder_etal}'s plot.
Other centrally bright remnants also exhibit small temperature
gradients and are near collisional ionization equilibrium
(or, as in the case of IC443
\citep{kawasaki_etal}, overionized).
The most relevant properties derived from the observations
of W44 are tabulated in Table~\ref{table:model}.

\clearpage

\begin{deluxetable}{ll}
\tablewidth{0pt}
\tablecaption{}
\tablehead{
\colhead{Physical Characteristic}
& \colhead{Previous Observational Support}}
\startdata
Age of $\sim$20,000 yr & Pulsar timing observations \\
Thermal X-ray Bright Center, no X-ray shell    & {\it{EXOSAT, ROSAT, ASCA}} \\
$\sim$ Isothermal X-ray Emission, $T \sim 4$ to $10 \times 10^6$~K & 
{\it{EXOSAT, ROSAT, ASCA}} \\
Partially formed cool shell & Radio continuum, \hone, \stwo, \halpha \\
Density enhancements in shell &  OH masers, \oone \\
\enddata
\label{table:model}
\end{deluxetable}
\noindent

\clearpage

\section{Suggested Explanations for W44's X-ray Characteristics}

Early researchers demonstrated that the observations are
inconsistent with the Sedov-Taylor model, the ``standard model'', 
in which
the SNR is in the adiabatic phase of development, and evolves in a 
homogeneous media without
thermal conduction, turbulent mixing, or evaporating clouds.
For example, the observations reveal no X-ray edge brightening, while the
adiabatic phase models predict a very bright edge.  The
observations reveal an X-ray bright projected center, while the adiabatic
phase models predict the center to be the dimmest region.
Also, the observations reveal near isothermality within the remnant,
while the adiabatic phase models predict strong temperature gradients.
To explain W44's very
bright center, dim edge, and near isothermality,
several scenarios have been proposed,
each with implications for the evolution
of supernova remnants and the physics of the interstellar medium.

\subsection{Obscuration and Shell Formation}

Because W44 is heavily obscured, the earliest explanation
for the lack of observed X-ray edge brightening was that the
lower energy X-rays from the anticipated edge were preferentially
absorbed by the intervening material.
\citet{long_etal} used Sedov models to rule out this explanation
for W28 and 3C400.2.
Subsequently, \citet{shelton_etal_95}, \citet{harrus_etal},
and \citet{shelton_etal} used
hydrodynamic codes to
show that the lack of a bright edge for W44 was more likely to be
because the remnant had evolved beyond the adiabatic phase and
into the radiative phase in which
the periphery has cooled (also see \citet{cox_etal}).
This interpretation is consistent with the radio continuum
evidence for shell formation and has been suggested for other centrally
bright remnants as well (i.e. 3C 397 \citep{safiharb_etal},
3C391 \citep{chen_slane}, IC443 \citep{kawasaki_etal}, N206 in
the Large Magellanic Cloud \citep{williams_etal}, and 0045-734
and 0057-7226 in the Small Magellanic Cloud \citep{yokogawa_etal})
While this evolutionary explanation could easily explain why
thermal composite remnants lack X-ray bright shells, it does
not additionally explain the bright emission from the center.  That
phenomena prompted the following theories.

\subsection{Evaporating Cloudlets}

Considering that the interstellar medium is inhomogeneous,
\citet{white_long} suggested that supernova shock fronts might
sweep past cloudlets.  If the cloudlets evaporate within a particular
range of timescales (similar to or longer than the age of the
SNR, but not long enough to provide significant
evaporated mass),
then they could load the remnant interiors
with material.  The advantages of this model are that the
entrained material could both increase the center's luminosity
and dampen the temperature gradient.  This model may also
explain \halpha, \oone, and OH maser observations
(see \citet{chevalier} for astrophysical analysis and observational
review).
The disadvantages are
that cloudlets have difficulty passing through the shock front intact
\citep{stone_norman}, cloudlet evaporation requires the unlikely
phenomenon of thermal conduction
through a sheath of dense and tangled magnetic fields lines,
the predicted remnant ages do not agree with other estimates
(compare ages calculated by \citet{rho_etal} and \citet{harrus_etal}
with \citet{wolszczan_etal_nodate}'s estimate for W44 or the
inconsistency in MSH 11-61A's age estimates \citep{jones_etal_98}),
and detailed
calculations do not predict the observed sharp X-ray surface brightness
gradient in the center (compare Figure 3 of \citet{harrus_etal} with 
Figure 13 of \citet{shelton_etal}).
Although no detailed hydrodynamic and spectral simulations
of supernova remnants with evaporating clouds have been published,
we can sketch the broad outlines of such spectra
from the characteristics of the evaporating cloud model.
Such spectra should have an evaporating cloud component and
a supernova remnant component.  The evaporating cloud model
claims that the majority of the flux from the central section of the remnant is
due to evaporating clouds, while the minority of the flux derives
from the supernova remnant itself.  The most recent evaporate should be
very underionized material embedded in a matrix of hotter gas, although
the older material would have had more time to approach equilibrium
with the hot matrix.
Later in the
paper, we will examine the
{\it{Chandra}} spectra for such features
(see Section 4.3).

\subsection{Entropy Mixing (such as Thermal Conduction or Bulk Motion)}

Thermal conduction or turbulent mixing (even if only operational
during the remnant's early evolution) could
redistribute entropy throughout the remnant's interior.
As a result, the interior could be made roughly isothermal,
isobaric, and isochoric (see calculations in \citet{cox_etal}
and hydrodynamic simulations in \citet{shelton_etal}).  In this scenario,
the remnant's interior, and especially its center,
would be denser than a non-conductive or unmixed SNR.
Because of its comparatively
larger interior density and more moderate temperature, the
entropy-mixed SNR would have a bright interior.
Thus, the advantages of this model are that it predicts a
luminous center and small temperature gradient.
The disadvantage is that it may require thermal conduction in
the presence of some magnetic fields, though in a far more likely
manner than thermal conduction across the tangled magnetic fields
surrounding the putative embedded clouds.
Further, as with the evaporating cloud
model, the detailed calculations
do not predict the sharp X-ray surface brightness gradient observed
near W44's projected center
(see Figure 13 of \citet{shelton_etal})
Note that two dimensional models of
thermally conductive remnants appear to require longer evolutionary times
or larger ambient densities in order for the remnant to evolve into
the radiative phase (compare 2-d models in \citet{shelton_etal} and
\citet{velazquez_etal} with 1-d models in \citet{shelton_etal}).
The spectra of an entropy-mixed SNR
should have a small temperature gradient with respect to radius.
With the exception of dust destruction effects,
the plasma should be nearing ionization equilibrium and may even
be recombining.  
This model will be tested by comparing
the {\it{Chandra}} spectra with detailed hydrodynamic and
spectral simulations in Section 4.3.

\subsection{Metal Enrichment Gradients}

Supernova explosions eject a few solar masses of metals into
the supernova remnant.  For example, \citet{thielemann_etal} predict
yields of 0.03 to 0.6 M$_\odot$ of neon,
0.012 to 0.2 M$_\odot$ of magnesium,
0.05 to 0.12 M$_\odot$ of silicon,
0.02 to 0.04 M$_\odot$ of sulfur, and
0.06 to 0.15 M$_\odot$ of iron for 13 to 25 M$_\odot$ progenitor stars.
If the yields were spread evenly throughout the remnant,
then they would make insignificant contributions to the total metallicity
of the plasma.  But, the material in a remnant of W44's age is not 
completely mixed from center to edge.
The ejecta should
preferentially enrich the remnant's center at the expense of
the remnant's periphery which predominately consists of swept up interstellar
matter.  Considering that the details of the mixing are not currently known,
let us suppose, 
that the ejecta are
mixed over one half of the remnant's radius.
Let us also suppose that the volume density in the center of the
remnant is about one tenth that of the ambient medium, as is found
for thermally conductive remnants entering the radiative phase of
evolution
\citep{cox_etal}.
If, in contrast, the entropy is not mixed, then the central volume
density would
be much less and the argument would become much stronger.
Assuming that the remnant is approximately 19pc $\times$ 16.5pc
$\times$ 23pc, the pre-explosion volume density was roughly 6~atoms~cm$^{-3}$
\citep{cox_etal,shelton_etal},
and the plasma has \citet{grevesse_anders} metal
abundances, then the central eighth of the remnant's volume
should contain
0.03 M$_\odot$ of neon,
0.01 M$_\odot$ of magnesium,
0.01 M$_\odot$ of silicon,
0.006 M$_\odot$ of sulfur,
and 0.03 M$_\odot$ of iron due to
the swept up interstellar matter.
Therefore the swept up ISM's contribution to  
the number of metal atoms in the central portion of the remnant
is smaller than the ejecta's contribution and therefore the 
swept up ISM metals should not
obscure a gradient in the ejecta's spatial distribution.

Another source of a metallicity gradient is dust destruction
within the remnant.
Dust in pre-shock ISM typically depletes
$95\%$ of the magnesium and silicon, $50\%$ of the sulfur,
and $98\%$ of the iron, but none of the neon from the gas phase
\citep{vancura_etal}.
These atoms will rejoin the gas phase as the dust is destroyed.
In time, they would experience collisions, raising their temperature
and ionization state.
We anticipate that the spatial pattern of
metal re-injection
is a complex result of multiple process.  On one hand,
the material in the remnant's center has experienced the strongest
shocks and the greatest lengths of time at the highest temperatures.
These factors favor dust destruction \citep{tielens_etal}
in the remnant's center.
On the other hand,
dust destruction behind non-radiative shocks (earlier in the
remnant's history) is a strong function of the volume density and
therefore of the swept up post-shock column density \citep{vancura_etal}.
Considering that the zone of swept up material behind the shock front
grows in column density as a remnant ages,
the outermost parts of W44 would
be favored regions for dust destruction.

A metallicity gradient, whether from dust destruction or ejecta,
would result in a gradient in the specific luminosity, which
would contribute to the observed surface brightness gradient.
Such an argument
has previously been made in other papers, including
\citet{shelton_etal}, and
\citet{yokogawa_etal}.
Elevated abundances have already been observed in a few older
remnants, including both
centrally peaked type remnants (\citet{yokogawa_etal} for the
total spectra of two centrally peaked SMC remnants) and
edge brightened remnants (\citet{miyata_tsunemi}
for the interior of the Cygnus Loop).
In the past,
it was not possible to test W44 for radial abundance trends because
the previous observations lacked sufficient spatial resolution.
Now, abundance gradient measurements are viable
using the {\it{Chandra}} data and may be of use in explaining
W44's morphology as well as constraining future mixing and dust destruction
calculations.
We will report on the results
in Sections 4.2 and 4.3.

\subsection{Density Gradients}

Recently, \citet{petruk} pointed out that if the
ambient medium has a large scale density gradient, then
the shells of adiabatic-phase remnants will have non-uniform
X-ray surface brightness.  The portion of the shell at the denser end
can produce more X-rays than the portion of the shell on the more diffuse
end.  If the remnant is viewed along the direction of the
density gradient, then its projected center can appear
to be brighter than its extremities.  Note that in this model,
the X-ray bright shell at the periphery of the projected remnant would
also be somewhat bright.  Analyses of previous X-ray data did not
find such an X-ray bright shell (\citet{rho_etal}, \citet{harrus_etal},
\citet{shelton_etal}).  The {\it{Chandra}} data has finer spatial
resolution and we will use it for a refined search for an X-ray shell
(see Section 4.1).

\section{{\it{Chandra}} Observations and Analysis}


W44 was observed by {\it{Chandra}} for 46 ksec on October 31, 2000.
Six ACIS chips (ACIS-S chips 1 through 5 and ACIS-I chip 3)
were active during the observation.
The data were calibrated using caldb 2.2.1.
Since CTI correction using the software of \citep{townsley_etal}
was only available for two of the five ACIS-S chips,
and applying CTI corrections to the S3 data did not change
spectra fit parameters,
we did not apply the CIT corrections in order 
to retain chip-to-chip consistency over all of the chips in use.
Periods containing soft background flares \citep{cxc_mmback}
were determined from the light curve and removed.
Point sources were detected using the CIAO {\it{wavedetect}} algorithm.
The S4 chip was de-striped using the algorithm developed by
\citet{houck} as this analysis was done before the destreak task 
was implemented in CIAO.
The redistribution matrix files (RMFs) and auxillary response files (ARFs)
for extended regions were calculated
using the CIAO {\it mkwarf} and {\it mkrmf} tasks.

Our determination of the instrumental and cosmic background began with
the front-illuminated chip I3 and the back-illuminated chip S1,
which sampled off-remnant portions of the sky and
have measured instrumental backgrounds.
The instrumental background can be extracted from
observations of the dark moon or calculated from event histogram data.
In either case, the measured instrumental background must be
scaled to the strength of the background during the observation
using the 10.5 $-$ 12.9~keV range where the {\it{Chandra}} 
response is minimal.
However, as we found significant differences in the shape in that
spectral range between our data and the event histogram data
then extant, but found that the dark moon data matched the W44 data
quite well, we used the dark moon data for the spectral shape
of the instrumental background.

The cosmic background in the direction of W44 consists of
very soft foreground emission due to the Local Hot Bubble
and absorbed background emission from the hard Galactic ridge.
The cosmic background was determined from
a spectrum derived from the off-SNR region of chip I3,
after scaling and removal of the instrumental background.
Chip I3's combined cosmic and instrumental background
compares well with that of chip S5, the only other
front-illuminated chip having some exposure to the
off-SNR sky.
Given the differences in responses among the chips,
we did not directly subtract the chip I3 background
from other spectra.
Instead, we fit a reasonable thermal model to the chip I3,
cosmic background, and then used that model, scaled by
the solid angle covered, to remove the cosmic background
from chips entirely covered by the SNR.
To this model was added a powerlaw (which was not 
convolved with the response function) in order to 
model the contribution from low-level background flares
to the instrumental backgrounds on the BI chips.

Chips S2, S3, and S4 viewed the center and interior of
the remnant, without viewing off-SNR regions.
We constructed a background spectrum for chips S2 and S4
from the cosmic background spectrum determined from 
the chip I3 off-SNR observations
and the instrumental background spectrum determined from
the chip S2 observations of the dark moon.
We constructed a background spectrum for chip S3,
a back-illuminated chip, 
determined the cosmic background spectrum
from the cosmic background spectrum determined from 
the chip I3 off-SNR observations
and the instrumental background spectrum determined from
the chip S3 observations of the dark moon.
In general, the cosmic background is poorly characterized,
primarily due to the low count rates,
but this is not a significant problem.
Figure~\ref{fig:final} shows that
the contribution by the cosmic background
spectrum is comparable to that of the instrumental background
for the 0.5-2.0 keV energy interval and
is small compared to the strength of the W44 spectrum
on the S2-S4 chips in the spectral region of interest.

\subsection{Images}

The ACIS-S chips were ``laid along'' the long
axis of the supernova remnant, such that Chips
S2, S3, and S4 viewed the center of the remnant, while
chips S1 and S5 viewed the southeast and northwest
edges of the remnant and the adjacent sky.
The pulsar synchrotron nebula lies
near the center of the remnant \citep{wolszczan_etal_nodate}
and so was positioned on chip S2.
The {\it{Chandra}} observations of the pulsar
wind nebula are analyzed in a separate paper \citep{petre_etal}.
Although the pulsar and its wind nebula emit sufficient numbers
of X-ray photons to be significantly detected during the
{\it{Chandra}} observation,
the pulsar and nebular photons do not hinder
the observation of the W44 supernova remnant.
The ACIS I3 chip viewed the
western edge of the remnant.
Figure~\ref{fig:xray_radio} shows
the positioning of the ACIS chips with respect to
the radio image.
These X-ray images were smoothed with a
$11.8$ arcsec HWHM Gaussian.

	The {\it{Chandra}} data beautifully and clearly show
that the emissive region extends to the periphery of the remnant,
but that
the radio outline and the X-ray outline may not be identical.  The
X-ray emissive region may extend slightly beyond the radio synchrotron
emissive region in the northwest and vice-verse in the southwest.
Both the radio and X-ray surface brightnesses fade gradually near the
periphery, so the apparent discrepancy could be due to differences in
emissivities and observational sensitivities.
As in previous X-ray observations of W44, no X-ray bright shell is
discernible in the images.  The greatest surface brightnesses
near the remnant's edge on Chips S1, S5, and I3 are
$2 \times 10^{-4}$ counts s$^{-1}$ arcsec$^{-2}$,
$4.5 \times 10^{-4}$ counts s$^{-1}$ arcsec$^{-2}$, and
$2 \times 10^{-4}$ counts s$^{-1}$ arcsec$^{-2}$.
The corresponding background rates are
$8 \times 10^{-5}$ counts s$^{-1}$ arcsec$^{-2}$ for Chip S1,
$6 \times 10^{-5}$ counts s$^{-1}$ arcsec$^{-2}$ for Chip S5, and
$6 \times 10^{-5}$ counts s$^{-1}$ arcsec$^{-2}$ for Chip I3.
The average intensity of the central 20 arcmin of the remnant is
two to three times stronger than that of the outer 10 arcmin
of the remnant.
Much of the emission in the central 20 arcmin
originates in several bright regions
strung along the long axis of the remnant.

\subsection{Spectra}

As an introduction, we present the integrated spectra
from individual chips before discussing
the spectra taken from smaller regions of the chips.
The prominent lines and complexes in the ACIS chip S2, S3, and
S4 spectra (see Figure~\ref{fig:totalspectrum}) are
\nenine\ ($\sim$ 910~eV),
\neten\ (1010~eV),
\mageleven\ ($\sim$ 1340~eV),
\magtwelve\ (1450~eV),
\sithirteen\ ($\sim$1850~eV),
\sifourteen\ (1980~eV), and
\sfifteen\ ($\sim$2450~eV).
The prominant emission lines of \nenine, \neten, \mageleven, \magtwelve,
\sithirteen, \sifourteen, and \sfifteen\
are expected to be strong in
$\sim 10^{6.6} $-$ 10^{7.2}$ K plasma \citep{raymond_smith},
if the plasma is assumed to be in
collisional ionizational equilibrium plasma (we will discuss
non-equilibrium models later). 
The higher energy features are expected to be stronger in
$\sim10^{7.2}$~K
gas, while the lower energy features are expected to be stronger
in $\sim10^{6.6}$ to $10^{7.0}$~K gas \citep{raymond_smith}.
The spectrum of the central region of the remnant
(chip S3) is harder and more influenced by $\sim$1850~eV
(\sithirteen) and $\sim$ 2450~eV (\sfifteen) emission
lines than the spectra from the surrounding
regions (chips S2 and S4).  The contrast suggests that the
center is a few times hotter than the adjacent regions.

The X-ray emission detected by chips S2, S3, and S4
lacks the \ironseventeen\ spectral feature at 1120~eV.
As a result, we expect little emission in the
\ironseventeen\ feature at 1020~eV and attribute the
observed $\sim1000$~eV feature solely to \neten\ (1010~eV).
The dearth of very highly ionized 
iron is corroborated by the following 
spectral modeling as well as that of
the {\it{ASCA}} data \citep{harrus_etal}.
When the Fe abundance is allowed to vary in the ``VNEI''
model (variable abundance non equilibrium ionization spectral
model using \citet{mazzotta_etal}'s ionization fractions)
in the XSPEC spectral fitting software, 
the model which best fits the chip S3 spectrum between
photon energies of 0.35 and 8.00 keV 
has an Fe abundance relative to the other elements of 
$0.04 \pm  0.01$.
In this model, the
temperature ($T)$ is $9.8 \pm 0.1\times 10^6$~K,
the ionization timescale ($\tau$) is 
$2.0^{+\infty}_{-1.7} \times 10^{13}$ s cm$^{-3}$
(hence the plasma is near ionization equilibrium),
and the absorbing column density ($N_H$) is 
$9.2 \pm 0.1 \times 10^{21}$ cm$^{-2}$.
The reduced chi-square parameter ($\chi^2_{\nu}$) is 5.0.
Figure~\ref{fig:S3fit} displays 
this spectral model and the S3 data, while Table~\ref{table:fits}
reiterates the model parameters.
Although the model spectrum fits the gross features of the spectrum,
it overpredicts some feature intensities while underpredicting others,
such as the $\sim$850 eV bump.
Because the strength of any given
feature depends on the plasma's temperature and extent of ionization, 
models composed of two spectra 
provide a better match to the assorted features of this spectrum.
One reason why such models are physically understandable
is because the line of sight 
intersects differing parts of the remnant and possibly
plasmas of differing qualities.
The best fitting two component model is presented in  
Table~\ref{table:fits}.  As is shown in Figure~\ref{fig:S3fit},
this model fills in the $\sim$850 eV bump
with emission from an iron rich, poorly ionized, hot plasma.
It is also possible
to fill in the bump with emission from an iron rich, poorly ionized, 
plasma of lower temperature (see Table~\ref{table:fits}).  
The common and important aspect of the second component is that
it contains poorly ionized iron.
In both cases, the second component provides only $\sim10\%$ of the
flux of 0.35 to 8.00 keV photons.  This component can be explained 
if much of the iron had been locked in dust grains when the
supernova shock swept through the plasma.  If the iron was
released from the grains as or after the plasma was shock heated, 
then its ionization would
lag that of other elements.
Note that the chip S3 data and, to a lesser extent, the S2 and S4 data
exhibited a slight energy shift.  The relative shift between S3 and
S2/S4 spectra
has been seen in other data sets, reported to the Chandra Science
Center, and removed using the fitting software; the shift in S2/S4 may
be due to the long-term gain drift for which this data had not been
corrected.

In comparison to the central region, the 
single temperature, non equilibrium fits to 
the neighboring regions (chips S2 and S4) prefer somewhat
lower temperatures, $6.9 \pm 0.1 \times 10^{6}$~K and
$7.0 \pm 0.1 \times 10^{6}$~K, respectively.
The ionization timescales for these models are
$9.3^{+\infty}_{-5.5} \times 10^{12}$ s cm$^{-3}$ and 
$2.8^{+\infty}_{-1.4} \times 10^{13}$ s cm$^{-3}$,
respectively (indicating, again, that the plasma is near collisional
ionizational equilibrium), 
the relative Fe abundances are
$0.37 \pm 0.02$ and $0.50 \pm 0.02$, respectively, and
the absorbing column densities are
$1.2 \pm 0.1 \times 10^{22}$ cm$^{-2}$ and
$1.4 \pm 0.1 \times 10^{22}$ cm$^{-2}$, respectively.
The reduced chi-square parameters are 5.2 and 4.1, respectively.
Two temperature fits were also performed.
These model parameters are presented in Table~\ref{table:fits}.
In accord with the results of the
{\it{ROSAT}} data analysis \citep{rho_etal,shelton_etal}, 
we find that the temperature 
profile is far too flat to be 
consistent with Sedov model predictions.

\clearpage

\begin{table}
\begin{tabular}{llccccc}
\hline \hline
Chip & Spectral & $T$        & $\tau$        & Relative {\bf{Iron}} 
& $N_H$         & $\chi^2_{\nu}$ \\
     & Model    & ($10^6$~K) & (s cm$^{-3}$) & Abundance	    
& (cm$^{-2}$)   &            \\ \hline
S3 & vnei & $9.8 \pm 0.1$  & $2.0^{+\infty}_{-1.7} \times 10^{13}$ & 
$0.04 \pm 0.01$ & $9.2 \pm 0.1 \times 10^{21}$ & 5.0 \\
'' & 2 temp vnei 
   & $8.8 \pm 0.1$ & $2.0^{+\infty}_{-1.7} \times 10^{13}$ & 
$ 0.07 \pm 0.01$ & $1.2 \pm 0.1 \times 10^{22}$ & 2.7 \\
  &  and     & $49.0 \pm 2.7$ & $ 4.6 \pm0.1 \times 10^{9}$ & 
$6.51 \pm 0.66$  & ''    &     \\
'' & 2 temp vnei 
   & $9.1 \pm 0.1$ & $2.0^{+\infty}_{-1.7} \times 10^{13}$ & 
$0.08 \pm 0.12$ & $1.3 \pm 0.1 \times 10^{22}$ & 3.6 \\
  &  and     & $7.7 \pm 0.5$ & $1.4 \pm 0.1 \times 10^{10}$ & 
$2.43   \pm 0.32 $ & ''    &     \\
S2 & vnei & $6.9 \pm 0.1 $ & $9.3^{+\infty}_{-5.5} \times 10^{12}$ & 
$0.37  \pm 0.02$ & $1.2 \pm 0.1 \times 10^{22}$ & 5.2 \\
'' & 2 temp vnei 
   & $3.5 \pm 0.1$ & $3.6^{+\infty}_{-0.1} \times 10^{13}$ & 
$0.58 \pm 0.04$ & $1.6 \pm 0.1 \times 10^{22}$ & 2.5 \\
  &  and     & $12.5 \pm 0.2$ & $1.6^{+\infty}_{-1.3} \times 10^{13}$ & 
$ 0.00 \pm 0.11 $  & ''    &     \\
S4 & vnei & $7.0 \pm 0.1 $ & $2.8^{+\infty}_{-1.4} \times 10^{13}$ & 
$0.50 \pm 0.02$ & $1.4 \pm 0.1 \times 10^{22}$ & 4.1 \\
'' & 2 temp vnei 
   & $3.3 \pm 0.1$ & $6.5^{+\infty}_{-3.1} \times 10^{12}$ & 
$0.83 \pm 0.06$ & $1.7 \pm 0.1 \times 10^{22}$ & 2.2 \\
  &  and     & $ 10.4 \pm 0.2$ & $3.0^{+\infty}_{-2.7} \times 10^{13}$ 
& $0.24 \pm 0.06$  & ''    &     \\
\hline
\label{table:fits}
\end{tabular}
\end{table}
\noindent

\clearpage

In order to examine the spectral trends on finer angular scales,
we subdivided the central third of the ACIS S field of view
into 30 segments along the long axis of the remnant
(see Figure~\ref{fig:regions}).
We then compared the spectra with spectral models.
Because the previous analysis found most of the emission
to be from nearly equilibrium plasma, equilibrium models were used.
The temperatures found by 
fitting the segments' spectra with single temperature,
equilibrium, variable abundance spectral models (the VMEKAL model on 
XSPEC \citep{mewe_etal,kaastra,liedahl_etal}, with Ne, Mg, S, Si,
and Fe abundances allowed to vary)
are plotted in Figure~\ref{fig:temp}. Figure~\ref{fig:temp} also
displays the reduced $\chi^2$ for the fits.
Again, the temperature profile is flatter than predicted by the
Sedov model.

As Figure~\ref{fig:abund} illustrates,
the abundances of
neon, silicon, sulfur, and iron peak near
the remnant's projected center.
The most logical sources of these abundance gradients are
the ejecta distribution and more advanced dust destruction in the remnant's
center.  We suspect that both phenomena are operating 
for the following reasons.
Because neon, an inert element, cannot be easily bound into
dust, the neon abundance gradient cannot be attributed to dust
destruction.  Therefore,
the existence of a gradient in the neon distribution
indicates the presence of ejecta.
The iron abundance distribution suggests that dust destruction is also
occuring.
The iron-rich component in the two component fits to
the S3 full-chip data 
indicates that
the iron atoms are less ionized than the other elements.  
The paucity of equilibrium ionization iron and the presence of underionized
iron
could be explained if
iron atoms had been released from dust grains
during the remnant's lifetime.  The elevated abundances near the
remnant's projected center would contribute to this region's 
elevated surface brightness.
%
The 800 to 900~eV feature in chip S3's spectrum
appears in the spectra for the 
outer $2/3$ of the chip S3 segments
(see Figure~\ref{fig:stackspectra}) 
and some parts of the flanking chips, but is inconspicuous in the
spectra of the remnant's projected center.
Presumably, this feature arises from L-shell iron in less ionized gas.
As was the case for the chip S3 spectral fitting, this feature
cannot be fit by equilibrium spectral
models for a wide range in temperatures without overpredicting
other emission features, but can be fit
by an additional, iron-rich, underionized spectral component.
%

The bright 'knots' (diameter $\sim$ 1 to 4 arcsec, 
brightness increase 
$\stackrel{>}{\sim} 1 \times 10^4$ counts cm$^{-2}$ s${-1}$ on chip S4) in the
central $\sim$20~arcmin of the remnant have long intrigued 
researchers.  
However, before the {\it{Chandra}} era, no X-ray telescopes  
had the sensitivity, spatial and spectral resolution necessary to isolate 
and study the knots.  
Figures~\ref{fig:xray_radio} and \ref{fig:regions} show that
{\it{Chandra}} has successfully resolved the knots; thus the knots
do not appear to be composed of smaller substructures (down to the
resolution limit of the telescope).
Here, we search for spectral differences between
the knot emission and the inter-knot emission in the chip S4 data.
%
We are focussing our efforts on the chip S4 data, because these
data allow us to distinguish between radial variation and knot versus
interknot variation.
For the analysis, 
chip S4's area was segmented, based on 
brightness, into 10 regions.  These regions have less than
1.34, 1.68, 2.12, 2.42, 2.75, 3.15, 3.56, 3.89, and
4.36 or greater than $4.36 \times 10^{-4}$ counts cm$^{-2}$ s$^{-1}$, 
respectively, in the 0.7 to 3.0 keV energy range.  Each has
$\sim$8800 counts in this range.  
The regions are outlined in Figure~\ref{fig:levl_regions}.  
The spectrum from each of the regions
was then fit with variable abundance, non equilibrium spectral
models (VNEI, with Ne, Mg, S, Si, and Fe allowed to vary), as well as
variable abundance, equilibrium spectral models (VMEKAL, 
with Ne, Mg, S, Si, and Fe allowed to vary).  
The ionization
parameter in the VNEI fits hovers around $2 \times 10^{13}$ s cm$^{-3}$,
indicating that the gas is near collisional ionizational equilibrium.
Independent of fitting method, we found little trend in temperature
%
%
(see Figures 10a and b)
We did, however, find a trend in the
best fit elemental abundances.  
Figures 10a and b
show that the best fitting models for the
brighter regions have greater
neon, silicon, sulfur, and iron abundances 
than the best fitting models of the dimmer parts
of the chip, though the trend is not uniform. 
The most obvious interpretations of these fitting results
are that more metals have been released from dust and pockets of ejecta 
in the bright knots than in the dimmer regions.  
Because increasing the Ne, Mg, S, Si, and Fe abundances
results in larger emitted fluxes of 0.7 to 3.0 keV photons, 
the abundance gradient contributes to the observed 
surface brightness variation.  However, it does not entirely
explain the observed surface brightness variation.
If the dimmer plasmas observed with parts of chip S4
were to have the Ne, Mg, S, Si, and Fe
abundances of the brighter plasmas observed on other parts of chip S4, 
they would still be dimmer.

\subsection{Comparisons with SNR Models}

In this section, we compare the {\it{Chandra}} data with models of W44.
Section 3 lists several suggested explanations for W44's X-ray morphology
(obscuration, shell formation, evaporating clouds, entropy mixing,
metal enrichment, and density gradients).  As stated in Section 3,
the first and last
explanations are ruled out by observations and/or modeling.  The
second explanation is well supported, but cannot fully explain W44's X-ray
morphology.
Therefore, we assume that the remnant has
evolved into the shell formation stage 
while examining the entropy mixing (see \citet{cox_etal},
\citet{shelton_etal}, or Section 3.3), 
evaporating clouds (see \citet{white_long}, \citet{long_etal}, or
Section 3.2) and
metal enrichment (\citet{shelton_etal}, or see Section 3.4) explanations.

To create our model of the entropy mixed (thermally conductive or
turbulently mixed)
supernova remnant, we used a detailed hydrodynamic computer simulation.
The simulation employs
saturated thermal conduction,
non-ionizational equilibrium cooling, ionization,
and recombination rates, and magnetic tension and is
described in \citet{shelton}.  The ambient
density (6.2 atoms cm$^{-3}$), explosion energy ($1.0 \times 10^{51}$ ergs),
effective magnetic field strength ($2.1 \mu$G), age (20,000 yrs),
and solar metallicities \citep{grevesse_anders}
are equal to those used in the \citet{cox_etal} and
\citet{shelton_etal} articles on W44 simulations.
This model meets the observational constraints on age, radius,
and degree of \hone\ shell formation.
Spectral predictions of this model (though not this particular
computer simulation)
were compared with previous observations in \citet{shelton_etal}.
Their thermally conductive simulated remnant was centrally peaked
in X-ray emission, significantly more so than the
otherwise equivalent nonconductive simulated remnant but, still
less than the observed morphology (see Section 3.3 and 
Figure 13 of \citet{shelton_etal}). Furthermore, their spectral
results for the supernova remnant 
satisfactorarily matched W44's
{\it{ROSAT}} and {\it{Einstein}} data.  
The $\chi^2_{\nu}$ for the combined fit was 2.0.
For the {\it{ROSAT}} comparison,
the spectrum of the entire remnant was used;  for the
{\it{Einstein}} comparison, the spectrum of the central 6'
(in diameter) was used.

The source of W44's controversy is its bright projected center, whose
photons were easily outnumbered by photons from other parts of the
remnant in the {\it{ROSAT}} analysis.
With {\it{Chandra}} data, it is now possible to compare the spectrum
of the remnant's center with those of simulated models.
The best fit between the central region
(using the eighth rectangular segment from the left in
Figure~\ref{fig:regions}, though nearby regions gave similar results) 
and the simulated 0.7 to 2.2 keV spectrum 
has a reduced $\chi^2$ of 1.5 and is displayed in Figure~\ref{fig:models}.
In the fitting process, the column density of absorbing material 
is allowed to vary.  The best fit case has
a column density of $2.1 \pm 0.1 \times 10^{22}$~cm$^{-2}$.

We also performed a simulation without thermal conduction.
The resulting model, with its hot ($\sim 6 \times 10^8$~K),
but dim center (as in Figures 11 and 13 of \citet{shelton_etal}),
underpredicts the $\sim1.4$ and $\sim1.9$~keV
features of the observed spectrum (see Figure~\ref{fig:models}).
As a result, the reduced $\chi^2$ for the fit is 5.1.

The evaporating clouds scenario was explored by \citet{white_long}
and \citet{long_etal},
who predicted the X-ray surface brightnesses
as a function of radius, but did not predict the spectra.
We cannot adequately predict the spectrum of a supernova 
remnant containing evaporating clouds
without extensively modifying and
testing our supernova remnant simulation package.
If we were to approximate the evaporating clouds spectrum
using a non equilibrium, variable iron abundance spectrum and
were to add it to the above non thermally conducting SNR simulation
to yield a spectrum for comparison with the observations, we would
only be able to constrain the clouds model by its need to bolster
the weak $\sim1.4$ and $\sim1.9$~keV features in the non thermally conducting
SNR spectrum.  Such a task is achievable, but of limited value.
For the interested reader, the parameters of a successful model are:
$T = 6.0 \times 10^8$~K (set), $\tau = 1.7 \times 10^{10}$~s~cm$^{-3}$,
relative Fe abundance = 0.32, $N_H = 0.97 \times 10^{22}$~cm$^{-2}$,
and $\chi^2_{\nu} = 2.1$.  In this case, the evaporating clouds produce
$70\%$ of the 0.7 to 2.2 keV flux, while the non thermally conductive
SNR produces the remaining $30\%$.
The spectrum is displayed in Figure~\ref{fig:models}.
Other combined models, with set temperatures as low as $1 \times 10^8$~K 
were tried and found to succeed similarly well.  Although the goodness of
these fits appear to add credence to these models, the 
extreme level of approximation used and the lack of
physical tests (such as age, radial surface brightness profile, etc.)
are serious detractors.

Earlier, we found a metallicity gradient across the
remnant as well as a metallicity gradient associated with
knot brightness.  Given that a plasma's specific luminosity
is an increasing function of its metallicity, these gradients
may be contributing to the observed brightness variation.
In order to estimate the extent of the contribution
to the radial surface brightness gradient, we compared surface
brightness profiles found with and without a metallicity
gradient.  
The solid histogram in 
Figure~13
traces the surface brightness profile of a model 
plasma having W44's metallicity gradient (for Ne, Mg, S, Si, and Fe),
while the dashed histogram traces the surface brightness profile of a 
model plasma having constant abundances of these elements
(equal to the averages of those observed for segments $7\_1$ through
$7\_14$, an $\sim$3 arcminute swath).  Both models have identical 
densities, temperatures, and normalizations as a function of
off-axis angle.  These parameters were found by 
fitting VMEKAL models to the observations.
Figure 13
demonstrates that the abundance gradient contributes substantially
to the observed surface brightness gradient.  However, it does
not explain all of the central brightness.  
If the abundances had been constant (as in the dashed histogram), 
the SNR would have exhibited a somewhat centrally enhanced X-ray morphology,
similar to that predicted with the entropy mixed SNR simulation
and some of those predicted with evaporating clouds calculations.

\section{Summary}

	The {\it{Chandra}} image of W44 reveals that the X-ray luminous
region extends to the periphery of the remnant outlined in radio
observations.  These data, like the {\it{ROSAT}} data, show that
there is no X-ray shell and that the
center's surface brightness is as much as several times stronger
than that nearer to the periphery.  The enhancement can be
described as a surface brightness plateau combined with several bright
knots within the remnant's central $\sim$10 arcminutes.

	The remnant exhibits a roughly monotonic, but weak
spectral gradient, such that the hardest spectra originate near
the projected center. By fitting model spectra to the data, we found that 
between the center and $\sim8$' ($\sim6$ pc) from the center, the plasma's
temperature drops by less than a factor of 2 and the elemental abundances
drop by a factor of several.
The elevated neon abundance suggests that a portion of the 
observed metal abundance gradient is due to progenitor ejecta.
Additional spectral fits indicate that the gas phase iron
atoms are less ionized than the
gas phase atoms of other elements.  This delayed ionization
suggests that iron was released from dust after the gas
was shock heated, making this the first
X-ray evidence for dust destruction within supernova remnants.
Several bright knots dot W44's face.
The plasma in these knots was found to have a similar temperature
and degree of ionization as the plasma in the surrounding regions, but
to have significantly greater gas phase abundances.  
If the knots had been evaporating clouds, we would not have expected
their thermal and ionizational properties to match those of the surrounding
regions.  If the knots are not evaporating clouds, then 
their astrophysical origins are even more mysterious than previously thought.


W44 is thought to have evolved into the radiative phase,
and therefore no longer has a hot, X-ray emitting shell.
Acting alone, this effect would leave the remnant's center relatively
dim and result in a quasi-edge brightened appearance,
not yet consistent with the observations
(compare dashed curve with symbols in 
Figure 13 of \citet{shelton_etal}).
Previous papers have shown that
obscuration plays a role, but not the one originally
assigned to it.  Obscuration preferentially
diminishes the flux of softer X-rays emitted at the larger radii, 
but the effect is not strong
enough to explain the remnant's centrally brightened appearance.

The unexpectedly bright centers of W44 and similar remnants
inspired the \citet{white_long} hypothesis for evaporating clouds
within these remnants.  W44's bright knots were 
assumed to be evaporating clouds or associations of 
evaporating clouds, and so 
taken as further evidence for cloud evaporation.
In the evaporating clouds hypothesis, the material evaporated off
of the clouds 
contributes substantially to the density of hot gas in the remnant's center
and boosts the center's X-ray luminosity.
But, the predicted radial gradient in surface brightness in the 
evaporating clouds models for W44 is far 
flatter than the observed gradient
and W44's predicted age is far less than
the observationally determined age
(see \citet{jones_etal, wolszczan_etal_nodate, harrus_etal,rho_etal}).
Furthermore, we learned from the {\it{Chandra}} analysis that the temperature
and ionization parameter do not strongly and significantly
vary from the knot regions to the interknot regions, suggesting
that the knots do not signify cloud evaporate.
Alternatively, if the hypothetical evaporating clouds were 
much smaller then the knots 
then it would be useful to
compare the {\it{Chandra}} spectra with a spectral predictions from
models.  Unfortunately,
there are no published spectral predictions for this scenario.
We mocked-up spectral models by combining a detailed hydrocode 
SNR simulation with non-equilibrium models of hot, recently
heated gas (representing the gas evaporated off of clouds).
We found the combined models to be good fits to the
{\it{Chandra}} spectrum ($\chi^2_{\nu} \stackrel{\sim}{=} 2.2$),
though the combined models are not particularly physical and the
fits are not particularly constraining.

The entropy mixed model was previously
shown to predict a bright center, though also with a far smoother radial
gradient than observed.  
The {\it{ROSAT}} spectrum for the entire remnant and the 
{\it{Einstein}} spectrum
for the central 6' (in diameter) were found to be well matched to
the spectra predicted from a detailed hydrocode model which included
entropy mixing in the form of thermal conduction
\citep{cox_etal,shelton_etal}.
Capitalizing on {\it{Chandra}}'s spatial resolution, in this
paper we examined the source of the controversy,
the remnant's bright center.
We compared the spectrum from the remnant's center with detailed
hydrocode model predictions, finding them to be a good match 
($\chi^2_{\nu} \stackrel{\sim}{=} 1.5$).  The balance of
evidence favors the entropy mixing model over the cloud evaporation
model and over the standard, adiabatic, Sedov-Taylor model 
($\chi^2_{\nu} \stackrel{\sim}{=} 5.1$), but none of these models
explains {\bf{all}} of the central flux.


The observed radial gradient in abundances and the correlation
between the knots and abundances were new discoveries made possible by the
{\it{Chandra}} data.  The additional metals in the remnant's projected
center raise the flux, making the remnant's projected center brighter
than outlying areas.  The abundance gradient is nearly strong enough 
to fill the gap in brightness 
between the observations and either the 
entropy mixing or the evaporating clouds model.

\noindent
Acknowledgements:

We acknowledge the financial support from CXC grant GO1-2057A.
RLS acknowledges discussions on evaporating cloud models with Knox
Long and mixing with Sally Oey.
KDK acknowledges discussions on the instrumental background
with P. Plucinsky, C. Grant, B. Biller, S. Wolk, K. Arnaud,
K. Gendreau, and M. Markevitch.

\pagebreak

\clearpage

\figcaption{The {\it{Chandra}} image (color) overlayed on the
radio frequency map (contours) of W44.
For the {\it{Chandra}} image, the energy bandwidth is 700 to 2600~eV
and the smoothing HWHM is $11.8$ arcsec.  For the radio map,
the wavelength is 20 cm and the beamsize is $15.8 \times 15.4$ arcsec$^2$.
The radio map is from \citet{frail_goss_etal}.
\label{fig:xray_radio}}



\figcaption{{\it Top:} Comparison of the W44 spectra
from chips S2 and S4 with the instrumental background
and the sum of the instrumental and cosmic backgrounds.
The instrumental background 
(labelled ``Dark Moon on S2'') is that for chip S2
derived from the dark moon data, while 
the cosmic background spectrum (labelled ``Obs. Fore/Back on I3'')
was derived from chip I3.
{\it Bottom:} W44's spectrum and the instrumental background on chip S3.
\label{fig:final}}

\figcaption{ 
Spectra extracted from ACIS S2, S3, and S4 data.
Left Panel:  Spectrum of the projected center of the remnant, taken
with the S3 chip.
Right Panel:  Spectra of the projected off-center interior, taken
with the S2 (solid curve) and S4 (dotted curve) chips.  In each
plot, the low surface brightness curve is the particle
background.  The particle background bump coinciding with the
\sithirteen\ feature in the chip S3 spectrum is associated with
Al K$\alpha$. 
\label{fig:totalspectrum}}

\figcaption{ 
Left Panel: Data from the S3 chip (crosses) and best
fit single temperature non equilibrium model with variable Fe abundance
(solid histogram).  The model fits the gross features of the spectrum,
but underpredicts the $\sim 850$, 
$\sim$1350 (\mageleven) and $\sim$1850~eV (\sithirteen)
features and overpredicts the $\sim$1450~eV (\magtwelve)
feature.
Right Panel:
Data from the S3 chip (crosses) and best fit two component
non equilibrium model with variable abundances of iron
(solid histogram).  This
model better predicts the $\sim850$~eV, $\sim1450$~eV, and $\sim1850$~eV
features at the expense of underpredicting the $\sim1000$~eV feature.
The parameters of these models can be found in the text and
Table~\ref{table:fits}.
\label{fig:S3fit}}

\figcaption{ 
The long axis of the remnant has been segmented
into 30 regions, which are outlined in this 
700 to 2600~eV image and used in subsequent analyses.
\label{fig:regions}}

\figcaption{Plots of estimated plasma temperature and
reduced $\chi^2$
as a function of angle from the center of the remnant
(center of chip S3).  
Angles increase to the
northwest and decrease to the southwest.
Each plasma temperature estimate and reduced $\chi^2$ value corresponds
to the best fitting MEKAL spectral model for the spectrum extracted
from the appropriate region of the Chandra data.  The region
widths vary.  The symbols are placed at the centers of the
regions.
\label{fig:temp}}

\figcaption{
Plots of estimated abundances relative to solar as a
function of angle from the center of the remnant (center of chip S3).  
The first, second, third, and fourth panels, respectively, display the
neon, silicon, sulfur, and iron abundances relative to solar
found from VMEKAL fits to the data.
\label{fig:abund}}

\figcaption{Spectra from every third segment on the S3 chip.
The region numbering begins with ``region 7,0'', the most southeastern
region and terminates with ``region 7,15'', the most northwestern region.
The observed spectra are noted with crosses.  
The overlayed curves, noted by solid histograms, 
are the best fit depleted iron MEKAL equilibrium model spectra.
Below each spectrum lies a linear plot of the residuals.
The ordinate range on the residual plots
is -0.04 to 0.04 counts cm$^{-2}$ s$^{-1}$ keV$^{-1}$.
The $\sim850$~eV features, which are more obvious in the off-center
spectra, are difficult to fit with equilibrium models.
\label{fig:stackspectra}}

\figcaption{ 
The Chip S4 area was segmented based on a brightness
criterion.
\label{fig:levl_regions}}

\figcaption{Plot of model parameters versus region number
for the VNEI fits to the flux-selected regions on chip S4.
(Region 0 is the dimmest region on chip S4, while Region 9 is the
brightest.)
The best fit abundances of Ne, S, Si, and Fe
have significant gradients, but the best fit temperatures and
ionization parameters (Tau) do not.}
\label{fig:kvnei}

\figcaption{Same as Figure~10, but for the
VMEKAL fits.
}
\label{fig:kvmekal}

\figcaption{Top Panel: {\it{Chandra}} spectrum from the remnant's
center (using the eighth rectangular segment from the left edge
of chip 7), marked with crosses
and spectrum from the thermally conductive (or entropy mixed) hydrocode model,
marked by a histogram.  The residuals are also shown.  The reduced
$\chi^2$ for the fit is 1.5.
Middle Panel: {\it{Chandra}} spectrum from the remnant's center,
marked with crosses, and spectrum from the non-conducting hydrocode model,
marked by a histogram.  The residuals are also shown.  The reduced
$\chi^2$ for the fit is 5.1.
Bottom Panel: {\it{Chandra}} spectrum from the remnant's center,
marked with crosses, and spectrum from the non-conducting hydrocode model
combined with a model for hot, recently ionized, evaporated material,
marked by a histogram.  The residuals are also shown.  The reduced
$\chi^2$ for the fit is 2.2.
\label{fig:models}}

\figcaption{Fluxes of 0.7 to 2.2 keV emission, as a function of 
off-axis angle for a model plasma
having the observed Ne, Mg, S, Si, and Fe gradients (solid histogram)
and a model plasma having constant Ne, Mg, S, Si, and Fe abundances
(dashed histogram).  All other parameters (density, temperature, and
normalization as a function of off-axis angle) are identical for these
two models and were found by fitting VMEKAL models to the data.
The contrast between these plots demonstrates that the observed
metallicity gradient has contributed signficantly to the surface
brightness gradient.}
\label{fig:flux}

\newpage
\plotone{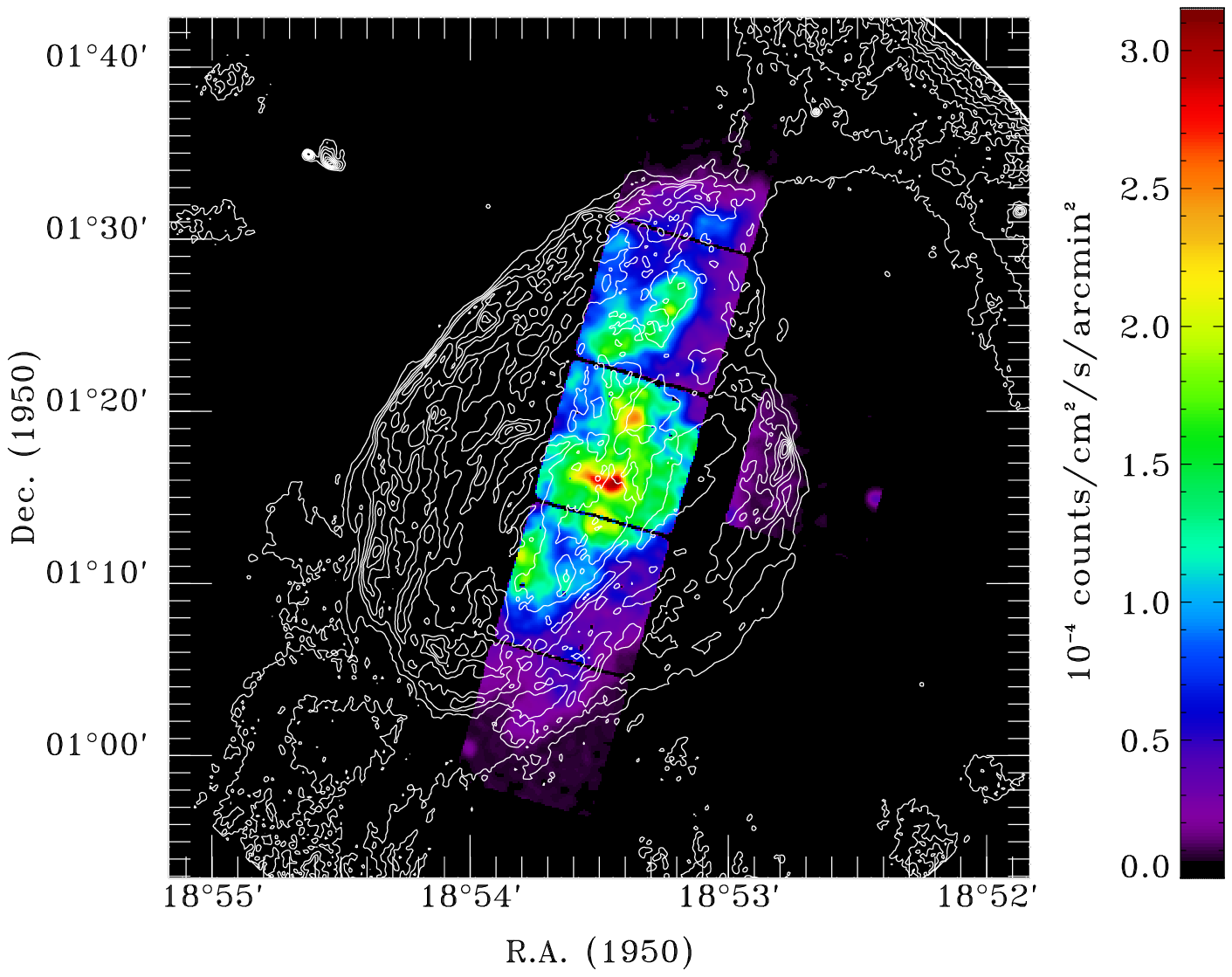}
\newpage
\plotone{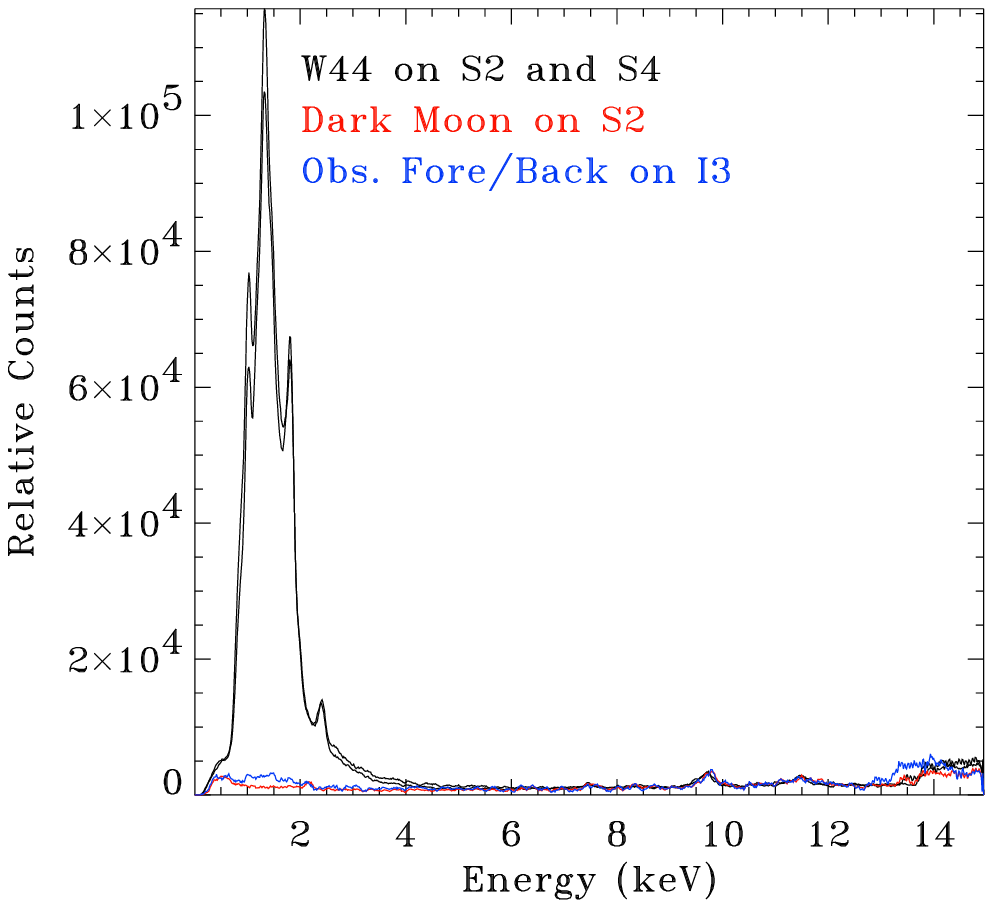}
\newpage
\plotone{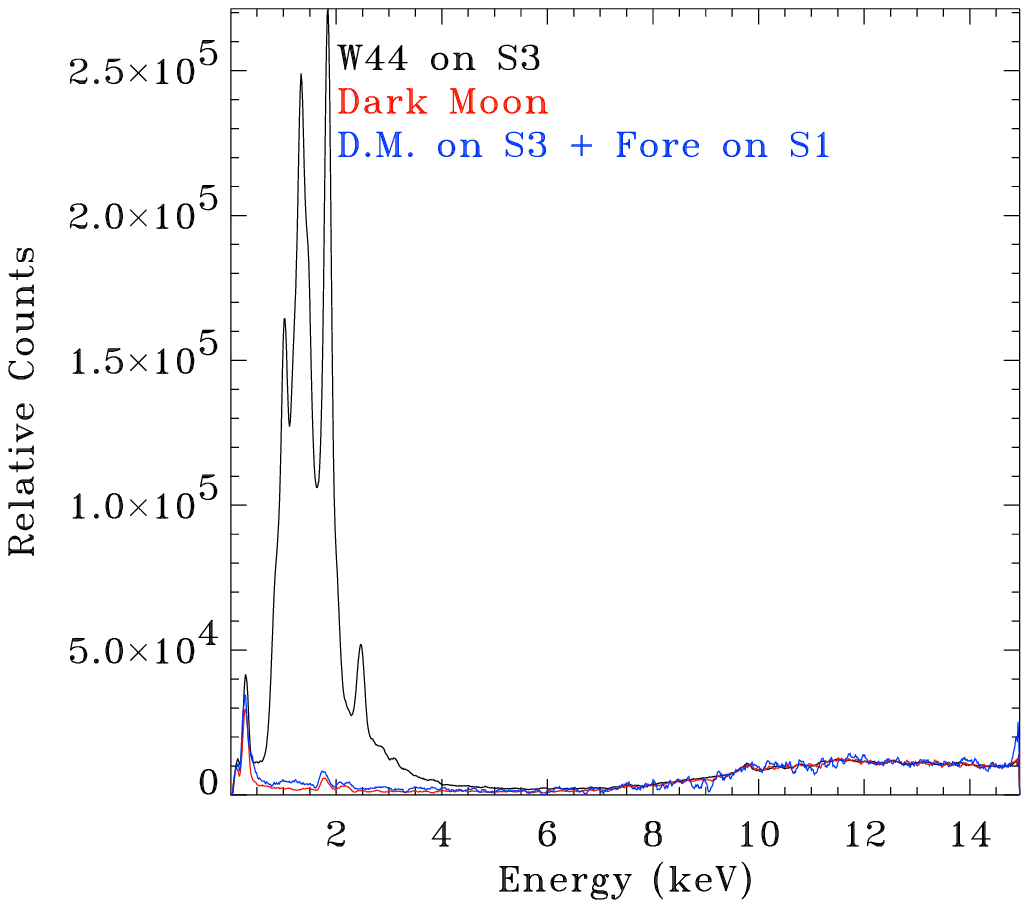}
\newpage
\plotone{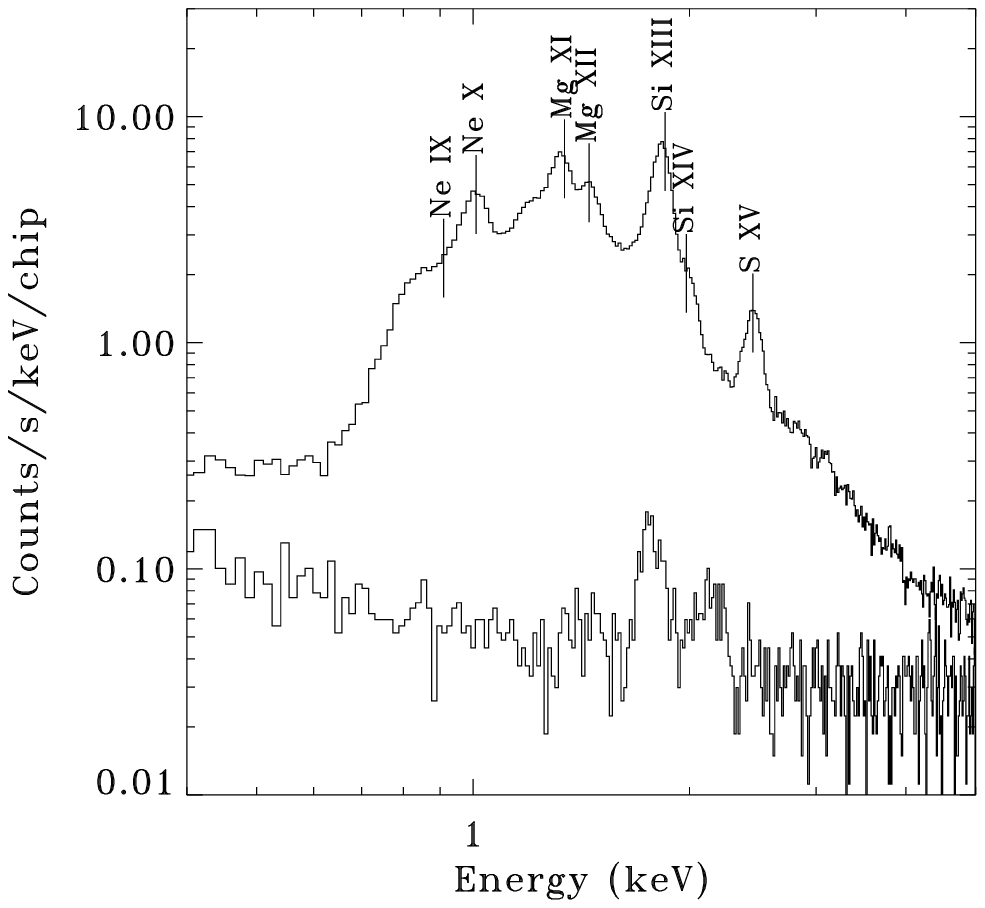}
\plotone{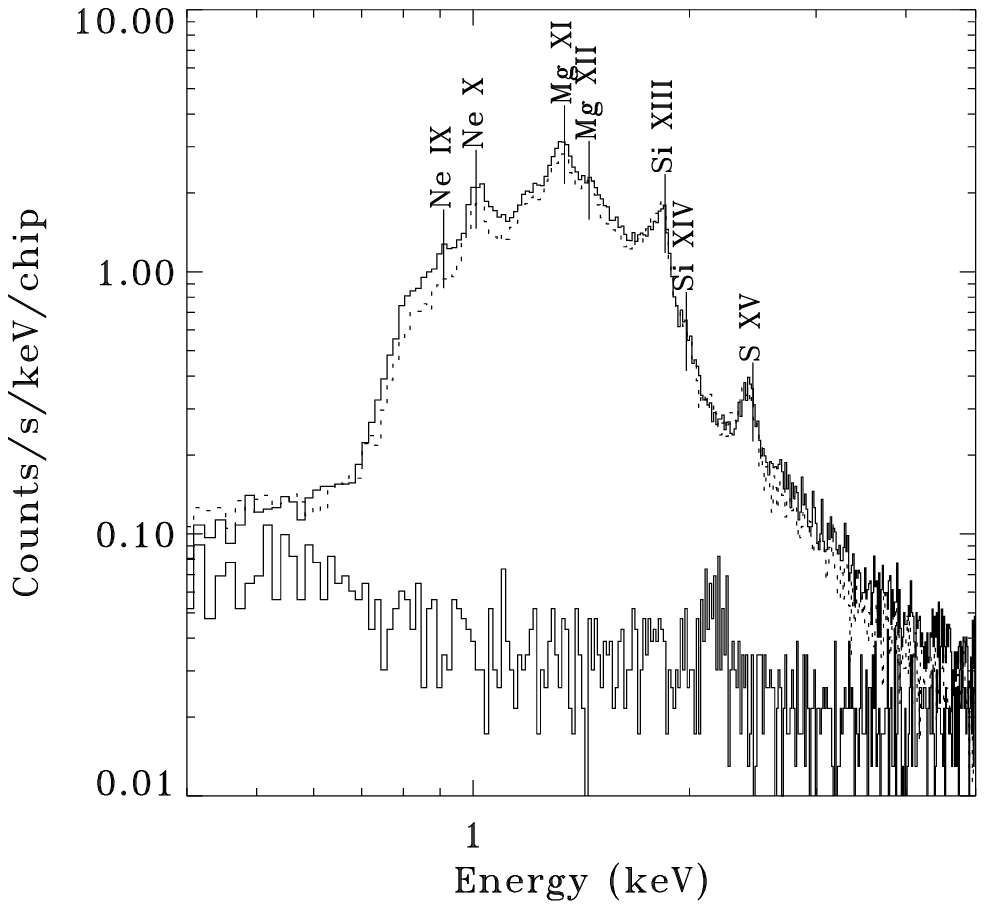}
\newpage
\plotone{figure4a.ps} 
\newpage
\plotone{figure4b.ps} 
\newpage
\plotone{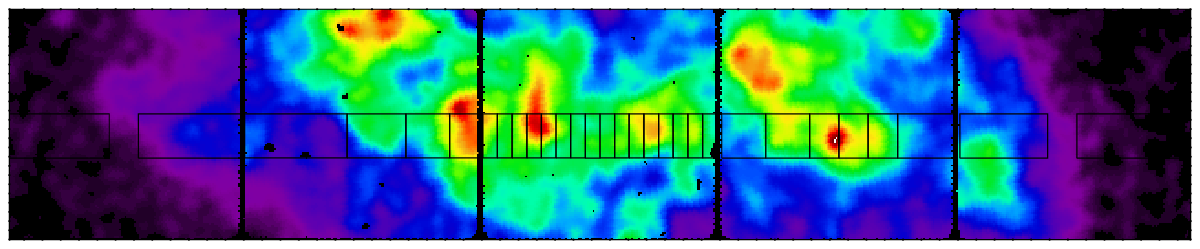}  
\newpage
\epsscale{0.5}
\plotone{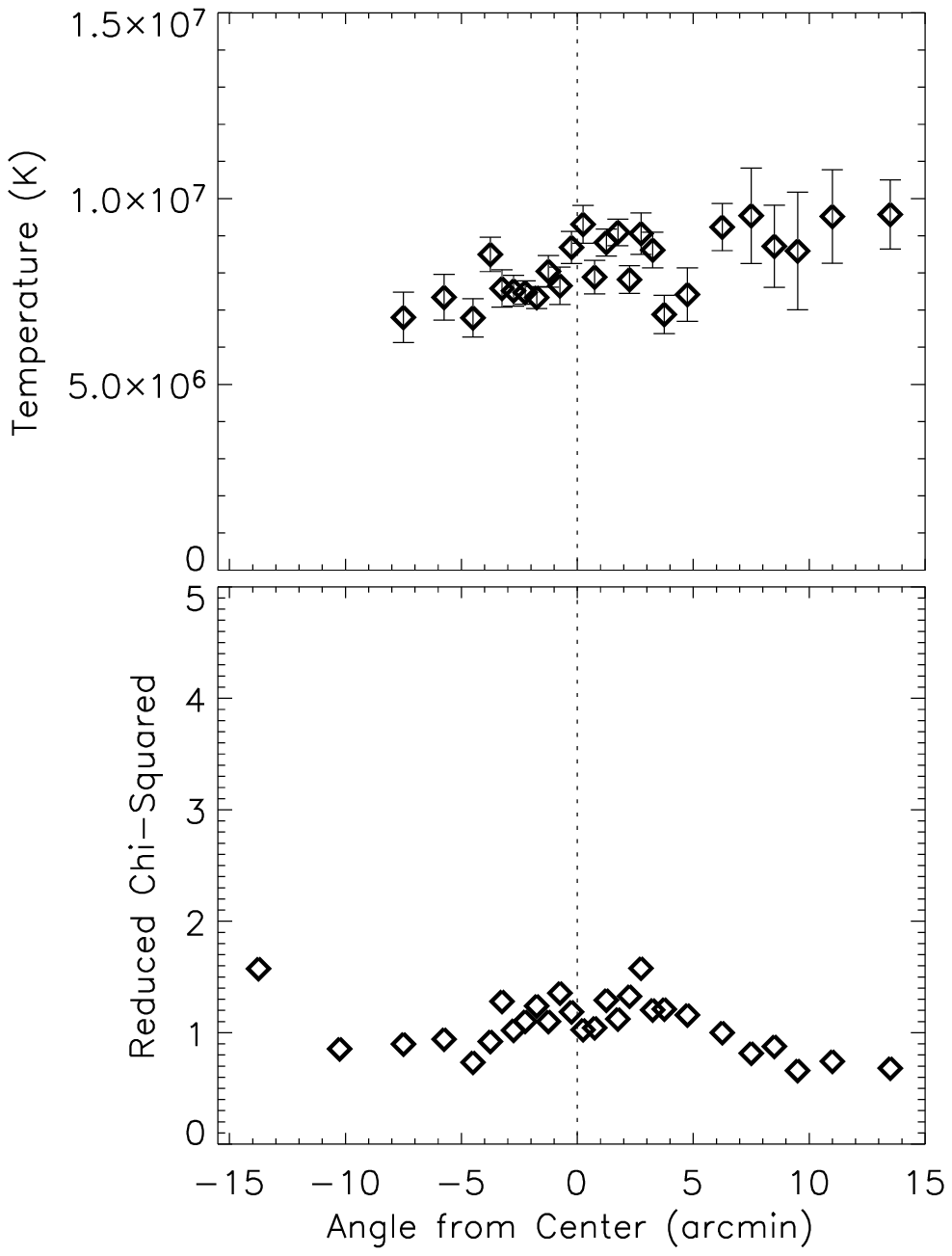}  
\newpage
\epsscale{0.5}
\plotone{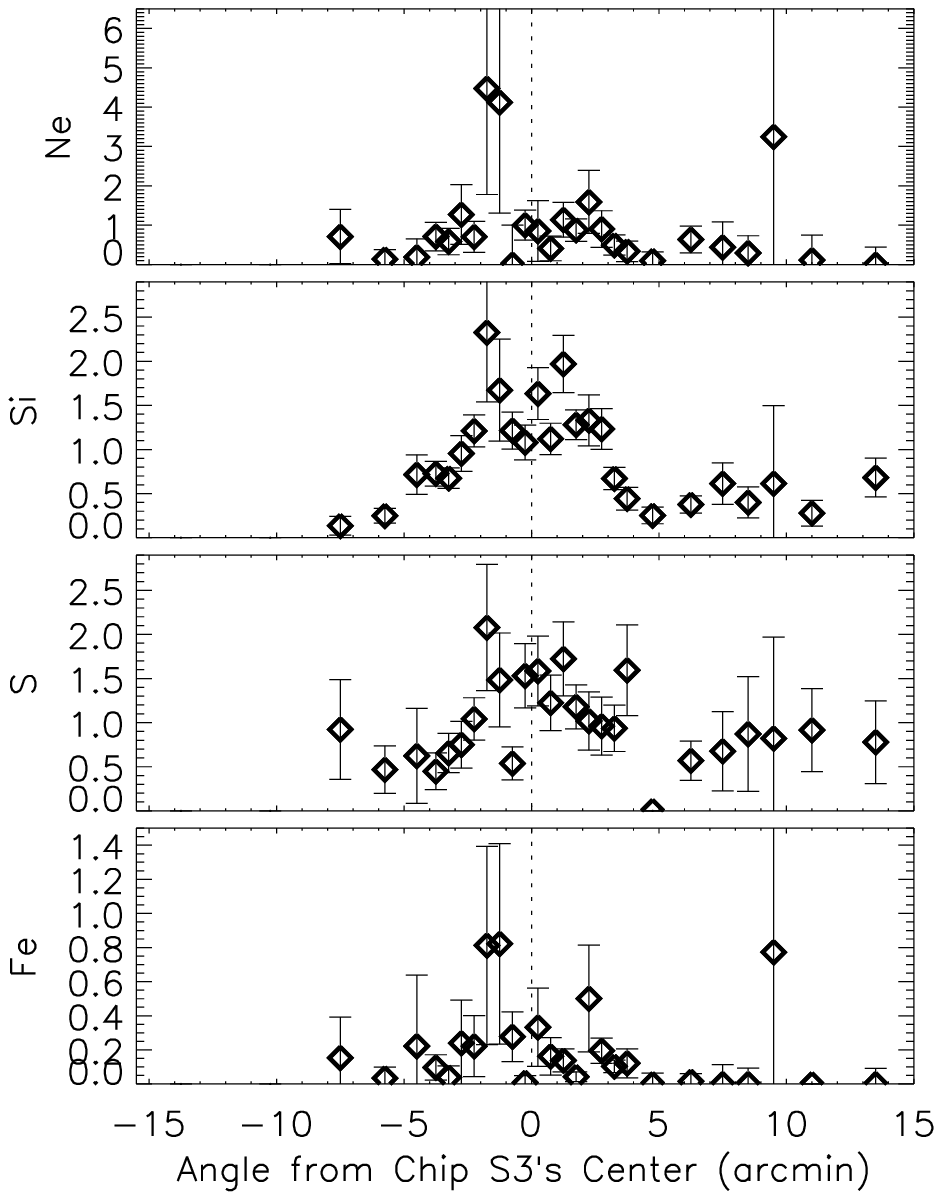}  
\newpage
\epsscale{0.5}
\plotone{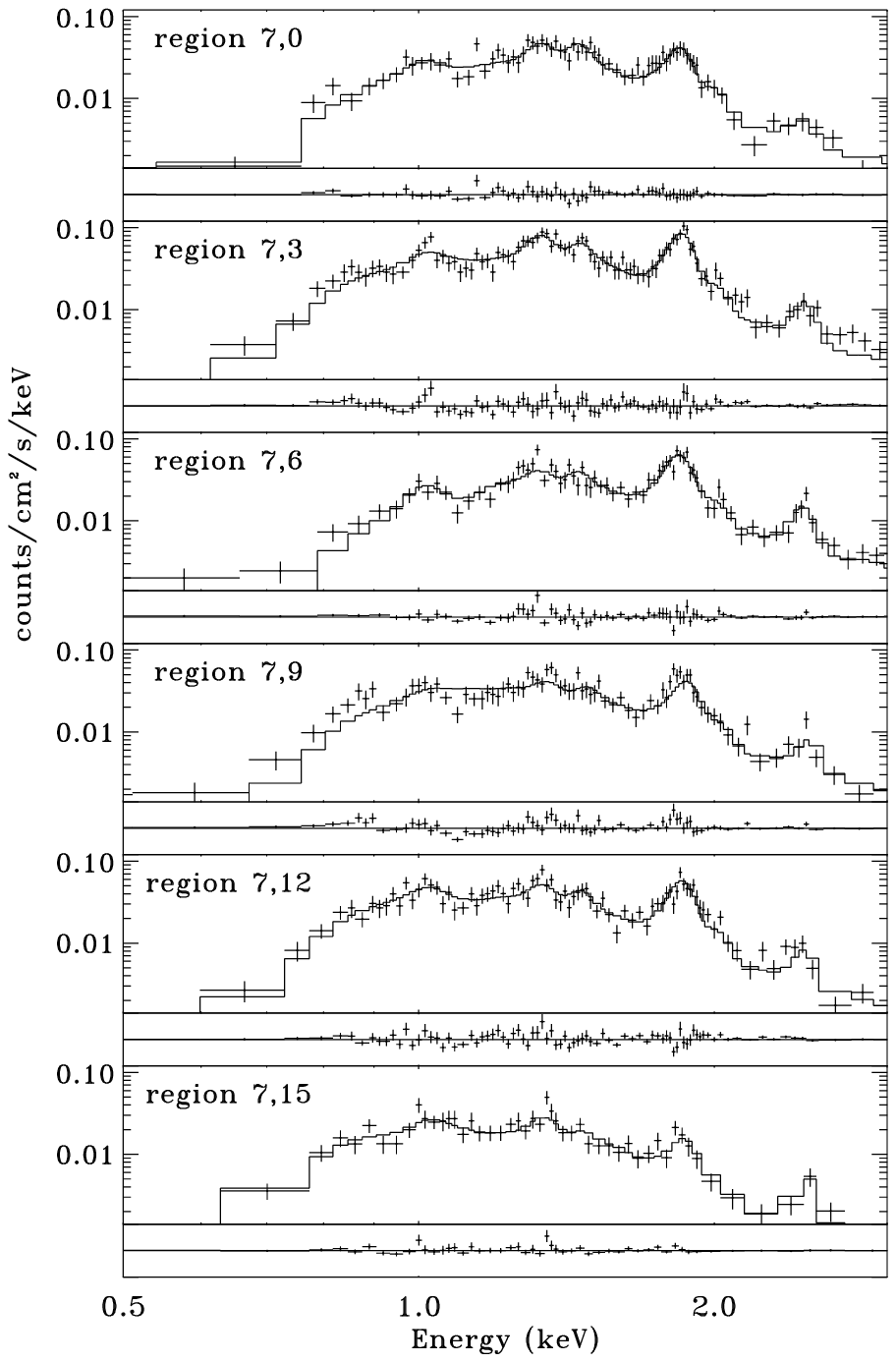} 
\newpage
\plotone{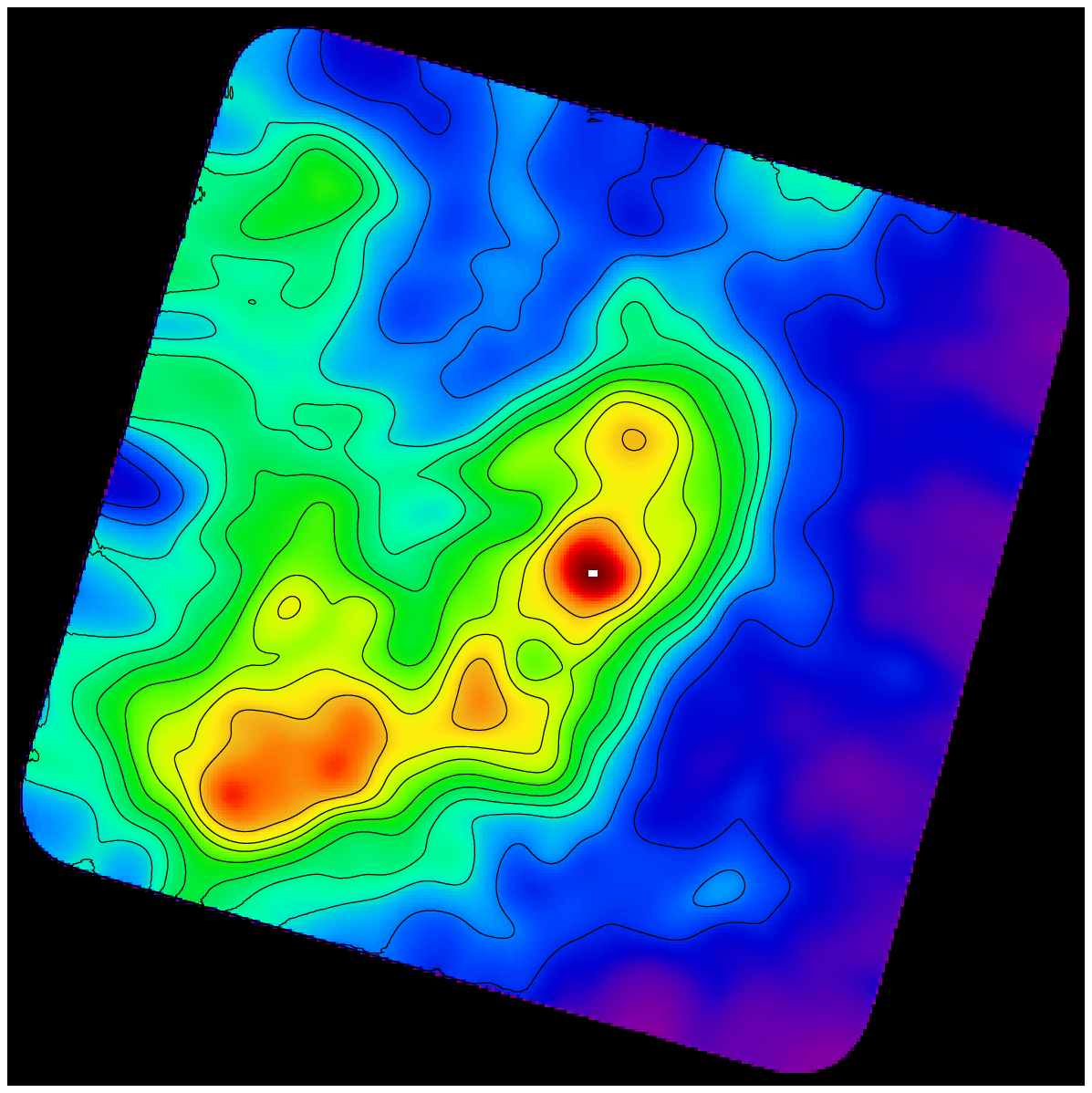}  
\newpage
\epsscale{0.5}
\plotone{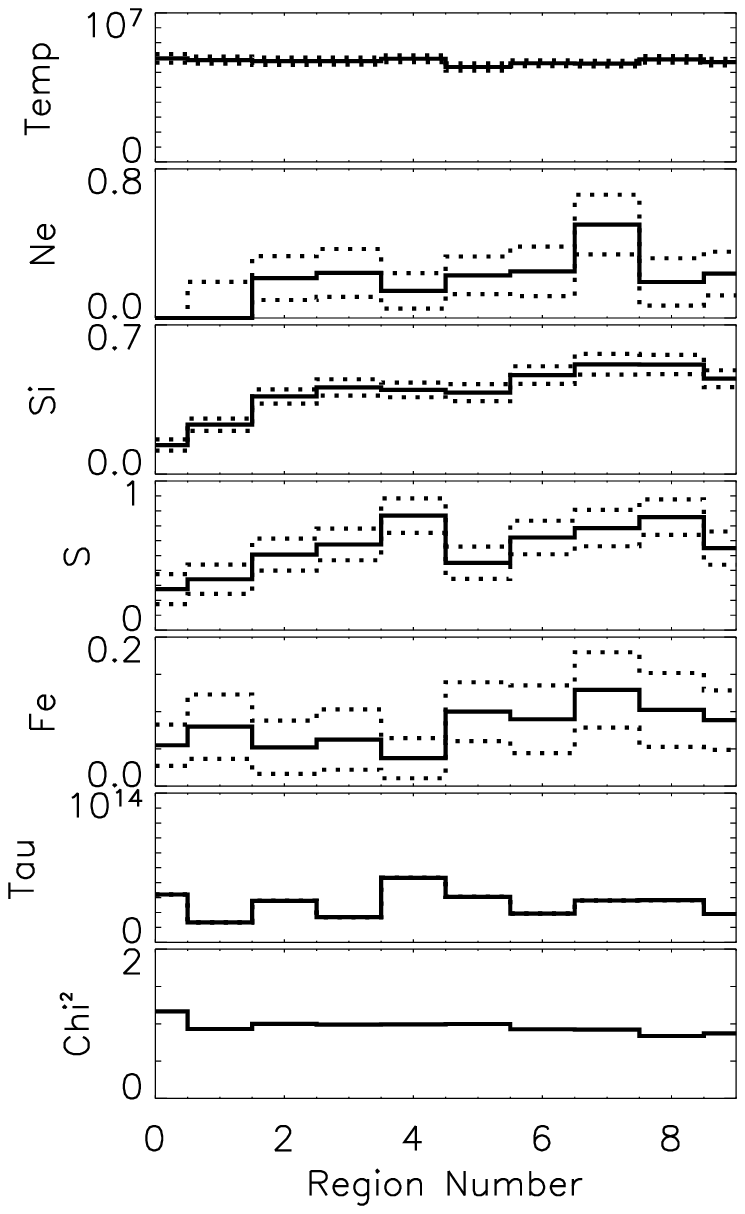} 
\epsscale{0.5}
\plotone{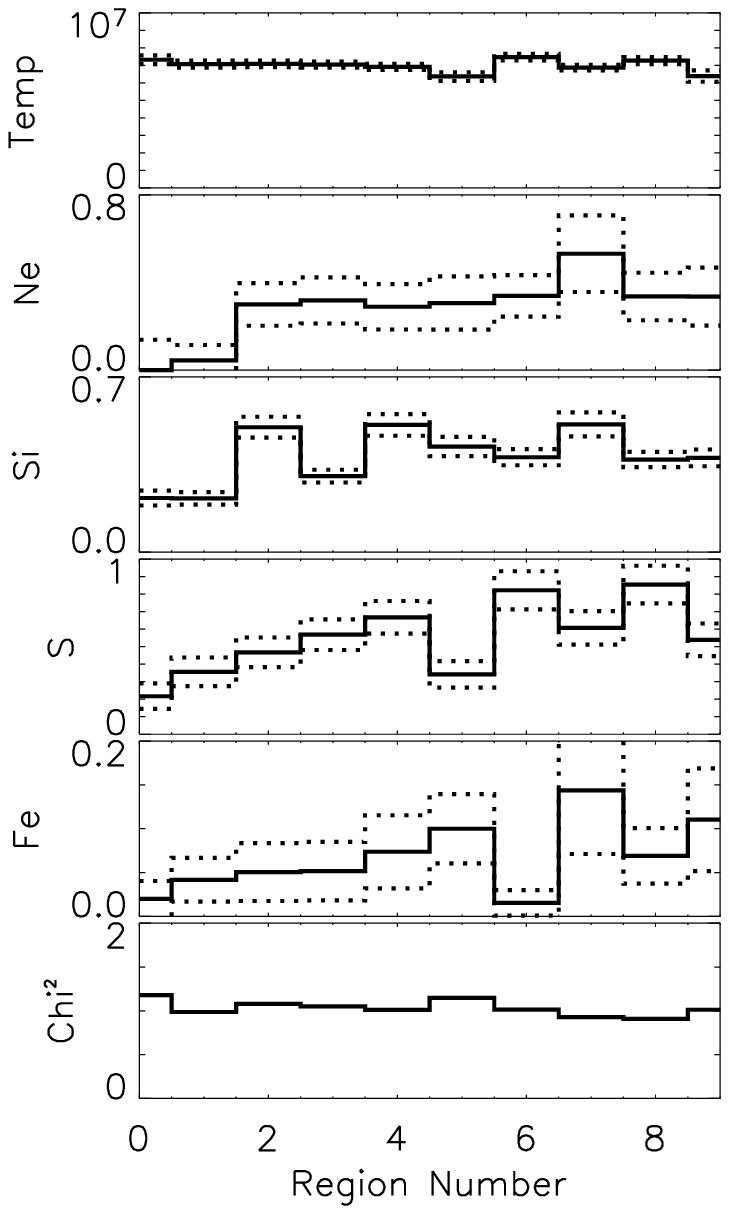} 
\newpage
\plotone{figure12a.ps} 
\newpage
\plotone{figure12b.ps} 
\newpage
\plotone{figure12c.ps} 
\newpage
\plotone{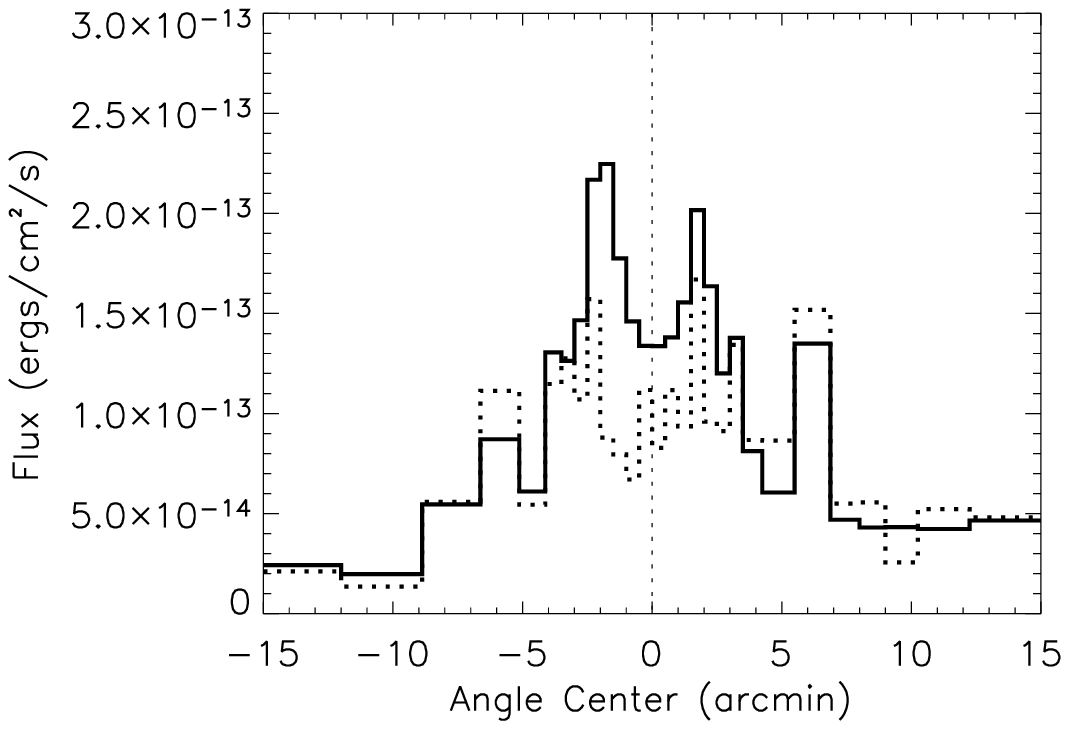} 

\end{document}